\documentclass[journal=jctcce]{achemso}
\usepackage{lmodern}
\usepackage{lmodern}
\usepackage[latin9]{inputenc}
\usepackage{amsmath}
\usepackage{amssymb}
\PassOptionsToPackage{version=3}{mhchem}
\usepackage{mhchem}

\makeatletter

\providecommand{\tabularnewline}{\\}

\@ifundefined{textcolor}{}
{%
 \definecolor{BLACK}{gray}{0}
 \definecolor{WHITE}{gray}{1}
 \definecolor{RED}{rgb}{1,0,0}
 \definecolor{GREEN}{rgb}{0,1,0}
 \definecolor{BLUE}{rgb}{0,0,1}
 \definecolor{CYAN}{cmyk}{1,0,0,0}
 \definecolor{MAGENTA}{cmyk}{0,1,0,0}
 \definecolor{YELLOW}{cmyk}{0,0,1,0}
}

\usepackage{xcolor}
\usepackage{pdfcolmk}
\usepackage{array}
\usepackage{multirow}
\usepackage{graphicx}
\usepackage{siunitx}
\usepackage[normalem]{ulem}

\newcommand{\wavenumber}{$\mathrm{cm^{-1}}$}

\makeatother

\title{Relativistic Complete Active
Space Self-consistent-Field Method with a Hierarchy of Exact Two-Component Hamiltonians}
\author{Xubo Wang}
\email{xwang405@jhu.edu}
\author{Sen Wang}
\author{Yixuan Wu}
\author{Lan Cheng}
\email{lcheng24@jhu.edu}

\affiliation{Department of Chemistry, The Johns Hopkins University, Baltimore,
MD, USA}
\begin{document}

\begin{abstract}

The development of a novel exact two-component (X2C) scheme with the inclusion of the picture-change correction
for the fluctuation potential, the ``X2Ccorr'' scheme, is reported, hereby establishing
a hierarchy of X2C schemes with systematic improvement for the treatments of relativistic two-electron contributions. 
Using benchmark X2C complete active space self-consistent-field (CASSCF) calculations for zero-field splittings in chalcogen diatomics,
the contributions of two-electron spin-orbit coupling, electron spin-spin coupling, and quantum electrodynamics
are carefully analyzed. 
The capability of the new Cholesky decomposition-based implementation for relativistic two-component
CASSCF method using super-configuration-interaction algorithms is further demonstrated with 
calculations for the low-lying electronic states of neodymium aqua-ions with up to the second coordination
shells.

\end{abstract}
\maketitle

\section{Introduction}

Special relativity plays an important role in chemistry and physics. \cite{Pyykko88,Dyall07,Reiher14}
The $s$- and $p$-electrons in a heavy atom move in a speed comparable to the speed of light, when they travel in the vicinity of a high Z nucleus. The relativistic increase in the masses for these electrons leads to radial contraction and energy stabilization. Note that even valence $s$- and $p$-electrons in a heavy atom can penetrate the core region and experience significant relativistic effects. Since the contraction of the $s$- and $p$-type orbitals improves the screening of the nuclear charge, the $d$- and $f$-electrons exhibit indirect relativistic radial expansion and energy destabilization. 
These relativistic kinematic effects have significant impacts on chemically relevant properties. 
Furthermore, special relativity introduces spin-dependent interactions, including spin-orbit coupling (SOC) and electron spin-spin coupling (SSC). They are responsible for important molecular spectral fine structures as well as spin-forbidden chemical processes. 
It is of importance to account for relativistic effects
to enable accurate computations of properties for molecules containing heavy atoms and of spin-dependent phenomena. 

The four-component Dirac-Coulomb-Breit (DCB) approach \cite{Sucher80,1996Saue,Dyall07,Sun21} in principle offers a rigorous framework for treating relativistic effects in atoms and molecules. The one-electron Dirac Hamiltonian provides rigorous treatments of one-electron scalar-relativistic and SOC effects. The instantaneous Coulomb interaction includes the spin-same-orbit (SSO) interaction of two-electron SOC (2e-SOC). The Gaunt term contains the spin-other-orbit (SOO) interaction of 2e-SOC and the SSC. On the other hand, the presence of the positronic degree of freedom in the relativistic four-component theory introduces significant computational overhead due to a large number of relativistic two-electron integrals. Therefore, relativistic two-component approaches \cite{Hess86,vanLenth93,Barysz01,Reiher06,Liu10,Saue11,Kutzelnigg12,Nakajima12} have been developed as cost-effective alternatives for molecular applications. In general, the two-component approaches decouple the electronic and positronic degrees of freedom in the four-component (effective) one-electron Hamiltonian 
to focus on electronic states. The exact two-component (X2C) theory \cite{Dyall97,Dyall01,Kutzelnigg05,Ilias07,Liu09,Peng12} 
that features a one-step decoupling for the matrix representation of the Dirac Hamiltonian 
has recently been established as the most promising two-component theory for molecular applications. 

The X2C theory in its one-electron variant (the X2C-1e scheme) \cite{Dyall01,Kutzelnigg05,Ilias07,Liu09} combines the X2C decoupling for the four-component one-electron Hamiltonian with the non-relativistic instantaneous two-electron Coulomb interaction. Since the one-electron term dominates the scalar-relativistic contribution, the spin-free version of the X2C-1e scheme (the SFX2C-1e scheme) \cite{Dyall01,Liu09,Cheng11b} has been established as a standard scalar-relativistic method. On the other hand, relativistic two-electron interactions make significant contributions to spin-dependent relativistic effects. \cite{Blume62,Blume63} It is essential to account for two-electron spin-dependent interactions in the X2C Hamiltonian, when aiming at accurate treatments of SOC and SSC. The central idea is to augment the four-component one-electron Hamiltonian with a mean-field approximation to the relativistic two-electron interactions. \cite{1996HessCPL} The X2C molecular mean-field (X2CMMF) approach \cite{Liu06,Peng07,Sikkema09} performs the X2C decoupling for the four-component Fock matrix. The X2CMMF Fock matrix has the same spectrum as the four-component Fock matrix. The underlying approximation of the X2CMMF scheme 
is the neglect of the picture change for the fluctuation potential. An X2CMMF coupled-cluster (CC) calculation has substantially improved efficiency for the integral transformation step compared to the corresponding four-component calculation. \cite{Sikkema09} On the other hand, it requires the solution of the four-component self-consistent-field (SCF) equations and hence the calculations of all relativistic two-electron integrals in each SCF iteration or the storage of these integrals. More efficient X2C model potential (X2CMP) schemes \cite{vanWullen05,Liu06,Peng07,Knecht22,Wang25a} have been developed to reduce the number of the molecular relativistic two-electron integrals required in the calculations. The X2CMP-DFT calculations use atomic densities as the model densities to eliminate the need to calculate molecular relativistic two-electron integrals. \cite{Liu06,Peng07}
The ``eamfX2C'' recipe \cite{Knecht22} for the construction of X2CMP Fock matrix uses atomic Hartree-Fock density matrices and requires the evaluation of a fraction of two- and three-center relativistic two-electron integrals only once. When treating the MP as a transferable potential, the X2CMP scheme has been shown to provide molecular properties that compare favorably with the four-component results. \cite{Wang25a} Furthermore, motivated by the locality of SOC, one-center approximation for relativistic two-electron integrals within the X2CMP framework leads to the X2C atomic mean-field (X2CAMF) scheme \cite{Liu18,Zhang22} and the closely related amfX2C scheme \cite{Knecht22}, which eliminate all molecular relativistic two-electron integrals. 

\begin{figure}[h]
\centering 
\includegraphics[width=6in]{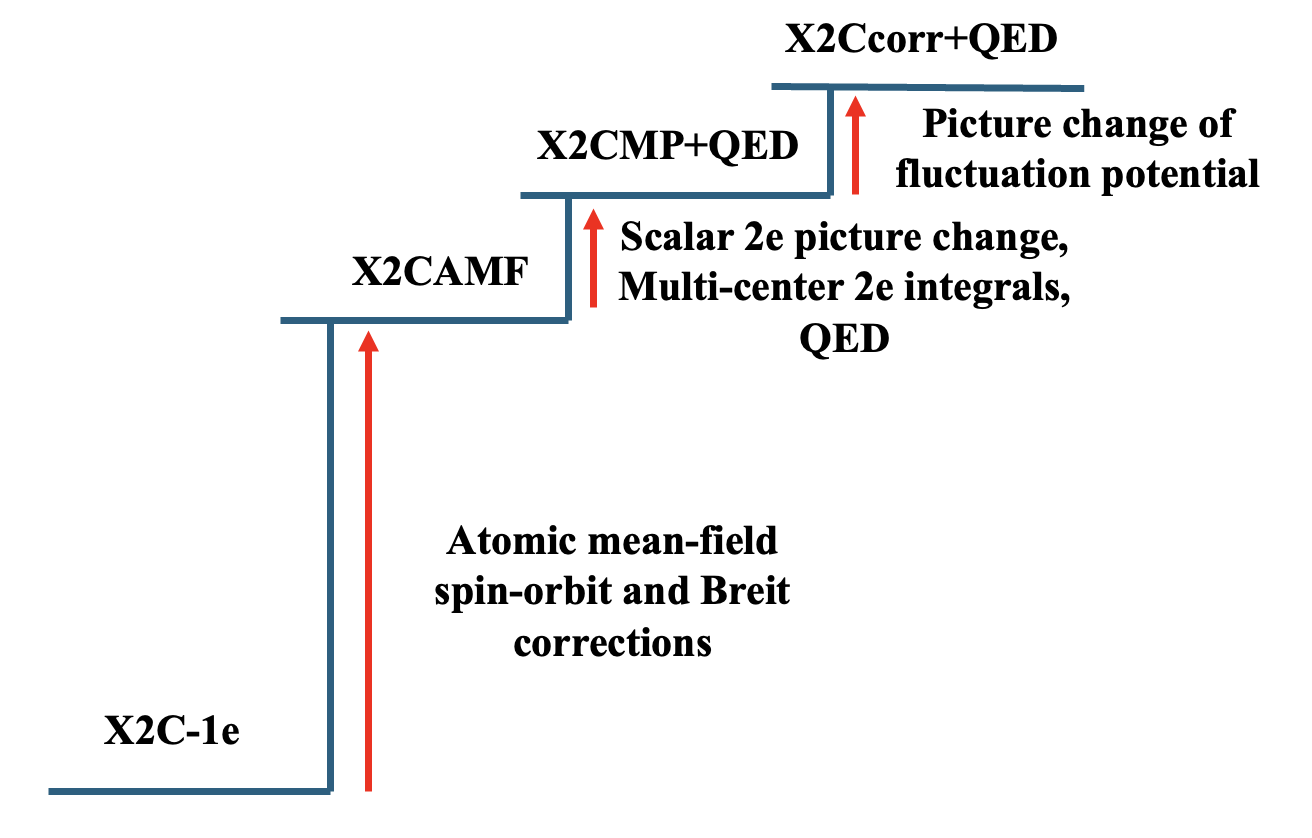}
\caption{A hierarchy of X2C schemes with systematically improved treatments of relativistic two-electron contributions.}
\label{fig:x2c}
\end{figure}

The X2CMMF and X2CMP schemes mentioned above typically use closed-shell or Kramers averaged density matrices to preserve the Kramers symmetry for the resulting Hamiltonian matrix. They cannot capture the contributions from the SSC at the mean-field level. 
In other words, when the molecular Hamiltonian is written in the occupation number representation and normal ordered with respect to a closed-shell reference function, the contribution from SSC resides entirely in the fluctuation potential. Consequently, the neglect of the picture change for the fluctuation potential in the X2CMMF and X2CMP schemes misses the contributions of the SSC. 
Note that the SSC makes significant contributions to important molecular parameters such as zero-field splittings. \cite{Vahtras02,Ganyushin06,Neese06}
In this paper, we develop an ``X2Ccorr'' scheme to address this challenge. 
The X2Ccorr scheme includes a partial integral transformation for the relativistic two-electron integrals within an active space, aiming to capture the leading relativistic correction to the fluctuation potential while maintaining the efficiency of X2C. As will be discussed in Section. \ref{X2C}, the X2C-1e, X2CAMF, X2CMP, and X2Ccorr schemes form a hierarchy of X2C schemes that systematically improve the treatments of relativistic two-electron contributions. While we focus on systematically improvable X2C schemes here, we mention that the screened X2C-1e schemes \cite{Filatov13,Ehrman23} provide simple treatments of two-electron spin-orbit interactions using screened nuclear charges in the one-electron spin-orbit term. \cite{Boettger00,Fedorov00}

We demonstrate the accuracy and usefulness of the X2C schemes using benchmark calculations for zero-field splittings of open-shell molecular species. The calculations of zero-field splittings in open-shell molecules necessitate treating several electronic states on the same footing and can be conveniently performed at the complete active space self-consistent-field (CASSCF) level. To enable CASSCF calculations in combination with the hierarchy of X2C schemes, the present work has developed a new implementation of relativistic spinor-based two-component CASSCF method in the PySCF program package. \cite{2018SunWCMS,2020SunJCP} 
Jensen {\it{et al.}} have formulated 
relativistic CASSCF method in the four-component framework
using the second-order norm-extended optimization
(NEO) algorithm in combination with Dirac-Coulomb (DC) Hamiltonian. \cite{1996Jo/rgenAa.JensenJCP}
Thyssen {\it{et al.}} have reported a corresponding implementation. \cite{Thyssen08}
By exploiting the spin separation of the DC Hamiltonian, \cite{Dyall94} Lipparini and Gauss have implemented the spin-free Dirac-Coulomb (SFDC) CASSCF method. \cite{2016LippariniJCTC} 
Shiozaki {\it{et al.}} have implemented the Dirac-Coulomb-Breit (DCB) CASSCF method in combination with
the density-fitting approximation to the two-electron integrals.\cite{2015BatesJCP,2018ReynoldsTJoCP}
Li {\it{et al.}}
have implemented the DCB-CASSCF method without introducing an approximation to
the two-electron integrals and have used 
density matrix renormalization group (DMRG) as an approximate
active space solver to enable the use of large active spaces.\cite{2023HoyerTJoCP} 
With the help of the density-fitting approximation and large-scale parallelization,
the implementation of four-component CASSCF method by Shiozaki {\it{et al.}}
has been demonstrated to be applicable to large molecules.
However, four-component CASSCF calculations are still significantly more expensive than
the corresponding two-component CASSCF calculations. 
Fleig \textit{et al.} have presented a two-component CASSCF formulation.\cite{1997Fleig}
Lee \textit{et al.} \cite{2003KimJCP,2013KimTJoCP} have implemented a two-component CASSCF method
using spin-orbit effective core potential (SOECP). 
Jenkins \textit{et al.} \cite{2019JenkinsJCTC}
and Evangelista \textit{et al.} \cite{Evangelista24}
have implemented two-component CASSCF methods using 
the screened X2C-1e Hamiltonians. 

The current X2C-CASSCF implementation aims to enable benchmark calculations for sizable molecules.
Therefore, we have used Cholesky decomposition (CD) of two-electron integrals \cite{Beebe77,Koch03, 2011AquilanteLTiCCaPMaA,2009PedersenTCA,Folkestad19} to improve the efficiency of the calculations
with controllable errors. 
Further, the current implementation employs the super-configuration-interaction (SCI) algorithm \cite{1971GreinCPL,1980RoosCP,1980RoosIJQC,1990MalmqvistJPC}
as an efficient alterative to the second-order algorithms \cite{1984Aa.JensenCPL,Werner85,2017SunCPL,2019KreplinEJ,2020KreplinJCP} for treating orbital rotation.
The details pertinent to the current implementation of X2C-CASSCF are presented in Section \ref{sec:theory}. 
We report the computational results for zero-field splittings in chalcogen
dimer molecules and for spin-orbit and ligand-field splittings in neodymium aqua-ions to demonstrate the accuracy of the X2C schemes and the capability
of the current two-component CASSCF implementation.
The computational details are summarized in Section \ref{sec:comp_details} and the computational results are discussed in Section \ref{sec:results}.
Finally,
we provide a summary and an outlook in Section \ref{sec:summary}.

\section{Theory}
\label{hamiltonians}

\subsection{A hierarchy of exact two-component (X2C) Hamiltonians} \label{X2C}

We first succinctly summarize the existing X2C schemes without the picture change for the fluctuation potential.
We refer the readers to the original work on these schemes \cite{Dyall97,Dyall01,Kutzelnigg05,Ilias07,Liu06,Peng07,Liu09,Sikkema09,Liu18,Zhang22,Knecht22,Wang25a} as well as recent reviews \cite{Liu24,Wang25a} for the details. 
Then we develop the X2Ccorr scheme to capture the picture change for the fluctuation potential.

\subsubsection{The X2C schemes without the picture change for the fluctuation potential} \label{X2C1} 

The ``no-pair'' four-component Dirac-Coulomb-Breit (DCB) Hamiltonian \cite{Sucher80} in the occupation number representation can be written as
\begin{eqnarray}
    \hat{H}^{\text{4c, no-pair}}=\sum_{pq}f^{\text{4c}}_{pq} \{a_p^\dag a_q\}
    +\frac{1}{4} \sum_{pqrs}g^{\text{4c}}_{pq,rs} \{a_p^\dag a_q^\dag a_s a_r\},
\end{eqnarray}
in which $\{\}$ represents normal ordering with respect to a reference function and
the summation runs over four-component positive energy states  $p, q, \cdots$. 
The Fock matrix element $f^{\text{4c}}_{pq}$
is the sum of the four-component one-electron matrix element $h^{\text{4c}}_{pq}$ 
and the mean-field two-electron contribution $g^{\text{4c,MF}}_{pq}$
\begin{eqnarray}
    f^{\text{4c}}_{pq} = h^{\text{4c}}_{pq}+g^{\text{4c,MF}}_{pq}~,~g^{\text{4c,MF}}_{pq}=\sum_i g^{\text{4c}}_{pi,qi}, \label{f4c}
\end{eqnarray}
in which $i$'s represent the occupied orbitals in the reference function. 
The antisymmetrized two-electron matrix element $g^{\text{4c}}_{pq,rs}$ in the DCB approach consists of contributions
from the instantaneous two-electron Coulomb interaction, the Gaunt term, and the gauge term. 

Similarly, an exact two-component (X2C) Hamiltonian in the occupation number representation is given by
\begin{eqnarray}
    \hat{H}^{\text{X2C}}&=&\sum_{pq}f^{\text{X2C}}_{pq} \{a_p^\dag a_q\}
    +\frac{1}{4} \sum_{pqrs}g^{\text{nr}}_{pq,rs} \{a_p^\dag a_q^\dag a_s a_r\}, \\
     f^{\text{X2C}}_{pq} &=& h^{\text{X2C}}_{pq}+g^{\text{nr,MF}}_{pq}~,~g^{\text{nr,MF}}_{pq}=\sum_i g^{\text{nr}}_{pi,qi},
\end{eqnarray}
in which $p, q, \cdots$ now represent the two-component orbitals
and $g^{\text{nr}}_{pq,rs}$'s are antisymmetrized matrix elements of the instantaneous two-electron Coulomb interaction. 
$h^{\text{X2C}}$ in the X2C-1e scheme, \cite{Dyall01,Kutzelnigg05,Ilias07} $h^{\text{X2C-1e}}$, is obtained by performing the X2C decoupling for $h^{\text{4c}}$. While $h^{\text{X2C-1e}}$ has the same spectrum as $h^{\text{4c}}$, the approximation of the X2C-1e scheme lies in the replacement of $g^{\text{4c}}$ with $g^{\text{nr}}$ in both the mean-field contribution and the fluctuation potential. This is often referred to as the neglect of the ``two-electron picture change''. 

The X2C molecular mean-field (X2CMMF) scheme \cite{Liu06,Peng07,Sikkema09} performs the X2C decoupling scheme for $f^{\text{4c}}$
to obtain $f^{\text{X2C}}$. The two-component Fock matrix in the X2CMMF scheme, hereafter referred to as ``$f^{\text{X2CMMF}}$'',
has the same spectrum as $f^{\text{4c}}$. Therefore, the X2CMMF scheme eliminates the 2e-pc error at the SCF level. 
Equivalently, we may rewrite $h^{\text{X2CMMF}}_{pq}$ as
\begin{eqnarray}
    h^{\text{X2CMMF}}_{pq}=f^{\text{X2CMMF}}_{pq}-g^{\text{nr,MF}}_{pq}, \label{mmfh}
\end{eqnarray}
which may be viewed as augmenting $h^{\text{X2C-1e}}$ with a correction for the 2e-pc contribution. 
The underlying approximation of the X2CMMF scheme is to replace $g^{\text{4c}}$ with $g^{\text{nr}}$ in the fluctuation potential.
In other words, the X2CMMF scheme neglects the picture-change correction for the fluctuation potential. 
The X2CMMF scheme has been shown to provide results in close agreement with the four-component results. 
However, we should emphasize that, when using a closed-shell reference function,
even the X2CMMF scheme completely misses the contribution from the electron spin-spin coupling (SSC). 

The X2CMMF scheme requires the evaluation of all four-component relativistic two-electron integrals in each SCF iteration or the storage of these integrals. To reduce the number of the required relativistic two-electron integrals, one may approximate Eq. (\ref{mmfh}) by replacing the molecular SCF reference function with a reference function built using atomic SCF solutions. This results in the X2C model potential (X2CMP) scheme. \cite{vanWullen05,Liu06,Knecht22,Wang25a} Specifically, the ``eamfX2C'' recipe \cite{Knecht22} uses the direct sum of atomic density matrices to construct $f^{\text{4c}}[\oplus \text{atoms}]$,
the X2C transformation of which produces $f^{\text{X2CMP}}_{pq}[\oplus \text{atoms}]$, and to construct $g^{\text{nr,MP}}[\oplus \text{atoms}]$
\begin{eqnarray}
    h^{\text{X2CMP}}_{pq}=f^{\text{X2CMP}}_{pq}[\oplus \text{atoms}]-g^{\text{nr,MP}}_{pq}[\oplus \text{atoms}]. \label{mph}
\end{eqnarray}
The X2CMP scheme requires a fraction of two- and three-center molecular relativistic two-electron integrals.
This significantly reduces the number of relativistic two-electron integrals compared to the X2CMMF scheme. When treating the correction to $h^{\text{X2C-1e}}$ as a transferable potential, the X2CMP scheme has been shown to provide accurate molecular properties. \cite{Wang25a}

An appealing scheme to further enhance the computational efficiency is to invoke a one-center approximation for the relativistic two-electron integrals. Invoking one-center approximation for relativistic two-electron integrals in Eq. \ref{mph} gives rise to the ``amfX2C'' scheme. \cite{Knecht22} Since the scalar-relativistic two-electron integrals are not as local as the two-electron spin-orbit coupling, the one-center approximation to these integrals leads to numerical instability of the amfX2C scheme. \cite{Liu24,Wang25a}
Although this numerical instability may be avoided by excluding the corrections for the diffuse basis functions, \cite{Wang25a} this requires further study and we will not use the amfX2C scheme in the present work. Instead, we will use the X2CAMF scheme \cite{Liu18,Zhang22} that 
explicitly excludes the scalar-relativistic two-electron integrals in the one-center approximation. 
The X2CAMF scheme performs a spin separation for the relativistic two-electron contributions to $g^{\text{4c,MP}}_{pq}[\oplus \text{atoms}]$ and excludes the contributions from the spin-free Dirac-Coulomb two-electron integrals. It then uses a one-center approximation for the rest of $g^{\text{4c,MP}}_{pq}[\oplus \text{atoms}]$, the X2C transformation of which gives rise to $\Delta g^{\text{X2C,SD}}_{pq}[\oplus \text{atoms}]$, including the spin-dependent (SD) terms in the Coulomb interaction together with the Gaunt term and the gauge term
\begin{eqnarray}
    h^{\text{X2CAMF}}_{pq}=h^{\text{X2C-1e}}_{pq}+\Delta g^{\text{X2C,SD}}_{pq}[\oplus \text{atoms}]. \label{amfh}
\end{eqnarray}
$\Delta g^{\text{X2C,SD}}_{pq}[\oplus \text{atoms}]$ is evaluated using atomic density matrices and atomic relativistic two-electron integrals. In other words, the X2CAMF scheme completely eliminates the molecular relativistic two-electron integrals and thus is computationally efficient. 

Compared to the X2CAMF scheme, the X2CMP scheme includes the scalar two-electron picture-change correction and the contributions from two- and three-center two-electron spin-orbit integrals. These contributions are relatively small. They have been found to be similar in magnitude to the contributions from quantum electrodynamics (QED) in calculations of core excitation energies. \cite{Southworth15,Wang25} Therefore, it appears to be necessary to include the QED contributions, when one aims to use the X2CMP scheme to improve the X2CAMF results. The leading QED corrections to molecular properties have been studied using Uehling potential for vacuum polarization (VP) \cite{uehlingPolarizationEffectsPositron1935,fullertonAccurateEfficientMethods1976} and local model potentials for self energy \cite{Pyykko03,flambaumRadiativePotentialCalculations2005,2010thierfelderQuantumElectrodynamicCorrections,2022SunagaJCP,2025JankeJCP}. Here we augment the $h^{\text{4c}}$ in the X2CMP scheme with vacuum polarization (VP) and self energy (SE) terms as formulated in Refs. 
\citenum{uehlingPolarizationEffectsPositron1935}, \citenum{fullertonAccurateEfficientMethods1976}, and \citenum{flambaumRadiativePotentialCalculations2005},
and hereafter refer to this scheme as ``X2CMP+QED''.



\subsubsection{The ``X2Ccorr'' scheme}

The X2C schemes, ranging from X2C-1e to X2CMP+QED schemes, as illustrated in Figure \ref{fig:x2c}, form a series of methods to systematically improve the treatments of relativistic two-electron contributions. The X2CAMF and X2CMP schemes use Kramers averaged atomic density matrices with no net spin polarization in the construction of the relativistic two-electron contributions. The mean-field contributions obtained using these Kramers averaged density matrices miss contributions from the electron spin-spin coupling (SSC). Even the X2CMMF scheme constructed using a closed-shell reference function does not include the contributions of the SSC. In calculations of zero-field splittings where the SSC often plays an important role, it is necessary to treat several closely lying electronic states on the same footing. 
Therefore, the inclusion of the picture-change correction for the fluctuation potential appears to be a plausible option for treating the SSC in X2C calculations. We hereby extend the X2C hierarchy by developing a new ``X2Ccorr'' scheme. The ``X2Ccorr'' scheme augments an X2CMMF Hamiltonian with a relativistic correction to the fluctuation potential within an active space
\begin{eqnarray}
    \hat{H}^{\text{X2Ccorr}}&=& \hat{H}^{\text{X2CMMF}}+\frac{1}{4} \sum_{tuvw} [g^{\text{4c}}_{tu,vw}-g^{\text{nr}}_{tu,vw}] \{a_t^\dag a_u^\dag a_w a_v\}, \label{HX2Ccorr} \\
    \hat{H}^{\text{X2CMMF}}&=&\sum_{pq}f^{\text{X2CMMF}}_{pq} \{a_p^\dag a_q\}
    +\frac{1}{4} \sum_{pqrs}g^{\text{nr}}_{pq,rs}\{a_p^\dag a_q^\dag a_s a_r\}.
\end{eqnarray}
in which $t, u, v, w$ refer to active space orbitals. Note that $f^{\text{X2CMMF}}$ is equivalent to $f^{\text{4c}}$ in the MO representation. 
In the present implementation, the X2CMP molecular orbital coefficients $C^{\text{X2CMP}}$ are back-transformed to the four-component picture to obtain the large- and small-component orbital coefficients, $C^{\text{L}}$ and $C^{\text{S}}$,
\begin{eqnarray}
C^{\text{L}}=RC^{\text{X2CMP}}~,~C^{\text{S}}=XC^{\text{L}},
\end{eqnarray}
using $X$ and $R$ matrices already constructed when transforming $f^{\text{4c}}[\oplus \text{atoms}]$ to $f^{\text{X2CMP}}_{pq}[\oplus \text{atoms}]$. 
These four-component coefficients are used to compute $g^{\text{4c}}$ and the four-component Fock matrix $f^{\text{4c}}$ as defined in Eq. (\ref{f4c}), in which the summation runs over the inactive orbitals in CASSCF calculations. 
The X2Ccorr scheme accounts for the picture-change correction for the fluctuation potential within the active space, which includes the contributions from SSO and SOO to the fluctuation potential as well as the SSC contributions.  

The construction of $\hat{H}^{\text{X2Ccorr}}$ described above requires one on-the-fly calculation of all relativistic two-electron integrals.
Since the active space of the CASSCF calculations presented here is small compared to the total orbital space, the $g^{\text{4c}}_{tu,vw}$ integral matrix can fit in the memory and the cost for the integral transformation is relatively small. The calculation of all relativistic two-electron integrals in the X2Ccorr scheme is a substantial overhead compared to the X2CMP scheme, since the X2CMP scheme only requires the evaluation of a fraction of two- and three-center relativistic two-electron integrals. On the other hand, this overhead is much smaller than the need to evaluate all relativistic two-electron integrals in each SCF iteration or to store these integrals in the four-component or X2CMMF scheme. Therefore, the X2Ccorr scheme presented here appears to be a computationally appealing approach to include the picture-change correction to the fluctuation potential. 

\subsection{Cholesky decomposition based relativistic two-component CASSCF method with super-CI algorithm}

\label{sec:theory} 

In the CASSCF method, the molecular orbitals (MOs)
are divided into three classes according to their occupation numbers, i.e.,
the inactive orbitals that are always occupied, the active orbitals with
no restrictions on their occupation, and the virtual orbitals that
are always unoccupied. In the following discussion, we use indices $i, j, k, l, \cdots$ to represent
for inactive orbitals, $t, u, v, w, \cdots$ for active orbitals, $a, b, c, d, \cdots$ for virtual orbitals, respectively. 
The general orbitals are denoted using $p, q, r, s, \cdots$.
A CASSCF wave function can be written as 
\begin{equation}
|\Psi^{\text{CASSCF}}\rangle=\hat{U}|\Psi_0\rangle~,~|\Psi_0\rangle=\sum_{I}C_{I}|\Phi_{I}\rangle.
\end{equation}
$|\Phi_{I}\rangle$'s 
are a set of orthonormal N-electron basis functions spanning a complete active space, 
which are chosen here as all Slater determinants obtained by distributing active electrons in an active space. 
$C_{I}$'s are the configuration interaction (CI) coefficients.

The unitary operator $\hat{U}$ accounts for orbital rotation and is parametrized as
an exponential of single excitations and de-excitations
\begin{equation}
\hat{U}=e^{\mathbf{\kappa}}~,~\mathbf{\kappa=}-\mathbf{\kappa^{\dagger}},
\end{equation}
in which
\begin{equation}
\mathbf{\kappa=} \sum_{pq} x_{pq} a^\dag_p a_q~,~ x_{pq}=-x_{qp}^\ast
\end{equation}
The CASSCF wave function is obtained by minimizing the CASSCF energy 
\begin{eqnarray}
    E^{\text{CASSCF}}=\langle \Psi^{\text{CASSCF}}|\hat{H}|\Psi^{\text{CASSCF}}\rangle
\end{eqnarray}
with respect to
the CI coefficients $C_{I}$'s and the orbital-rotation parameters $x_{pq}$'s. 
In the two-step procedure that we employ in this work, the update of $C_{I}$'s and $x_{pq}$'s 
are carried out alternately until convergence.
In each iteration, the orbital-rotation parameters $x_{pq}$'s are determined by minimizing $E^{\text{CASSCF}}$
with a fixed set of CI coefficients. After this, a set of new CI coefficients 
are obtained by solving the full CI (FCI) equations in the active space with the new orbitals.
We focus the following discussion on the determination of the orbital-rotation parameters
using the super configuration interaction (SCI) scheme and the implementation using Cholesky-decomposed two-electron integrals.

\subsubsection{The SCI scheme for two-component CASSCF}


In the SCI scheme, the action of the orbital-rotation operator is linearized
and this gives rise to a CI singles (CIS) form of the CASSCF wave function
\begin{eqnarray}
|\Psi^{\text{CASSCF}}\rangle &\approx& |\Psi^{\text{CASSCF}}_{\text{SCI}}\rangle = |\Psi_0\rangle+\sum_{pq} x_{pq}|\Psi_q^p\rangle, 
\end{eqnarray}
in which \(|\Psi^p_q\rangle\equiv a_p^\dagger a_q|\Psi_0\rangle\) denotes a singly excited basis function. 
The orbital rotation within the active space is redundant in the minimization of CASSCF energy, since a FCI is performed within the active space. Therefore, the single excitations in the active space, $x_{tu}a^\dag_t a_u$, are not included in the parametrization. 
Among the rest of the single excitations,
only $a^\dag_a a_i$, $a^\dag_u a_i$, and $a^\dag_a a_u$ make nonzero contributions when acting on $|\Psi_0\rangle$.
$|\Psi^{\text{CASSCF}}_{\text{SCI}}\rangle$ thus can be written as
\begin{eqnarray}
|\Psi^{\text{CASSCF}}_{\text{SCI}}\rangle &=& |\Psi_0\rangle+\sum_{ai} x_{ai}|\Psi_i^a \rangle+\sum_{ui} x_{ui}|\Psi_i^u\rangle+\sum_{au} x_{au}|\Psi_u^a\rangle, \label{sciwav}
\end{eqnarray}
Taking the expectation value of the Hamiltonian operator using the approximate CASSCF wave function in Eq. (\ref{sciwav}), one obtains an approximate second-order energy functional 
\begin{equation}
E^{\text{CASSCF,(2)}}_{\text{SCI}}(\mathbf{x})=E_0+\mathbf{g^\dagger} \mathbf{x}+\mathbf{x^\dagger} \mathbf{g}+\mathbf{x^\dagger \mathcal{H}x},
\end{equation} 
in which $E_0$ is the energy of the trial CASSCF wave function $|\Psi_0\rangle$
\begin{eqnarray}
    E_0=\langle \Psi_0|\hat{H}|\Psi_0\rangle,
\end{eqnarray}
\(\mathbf{g}\) contains the Hamiltonian matrix elements between \( |\Psi_0\rangle\) and the singly excited basis functions \(|\Psi^p_q\rangle\) 
\begin{equation}
    g_{pq} = \langle \Psi^p_q|\hat{H}|\Psi_0\rangle, 
\end{equation}
and the matrix elements of $\mathcal{H}$ are given by 
\begin{equation}
     \mathcal{H}_{pq,rs}=\langle \Psi_q^p|\hat H|\Psi^r_s\rangle-\langle \Psi_0|\hat{H}|\Psi_0\rangle\langle \Psi_q^p|\Psi^r_s\rangle.
\end{equation}
Stationary conditions of this energy functional, $\frac{\partial E^{\text{CASSCF,(2)}}_{\text{SCI}}(\mathbf{x})}{\partial x^\ast_{pq}}$, lead to a generalized eigenvalue equation
\begin{equation}
    \begin{pmatrix}
        0 & \mathbf{g}^\dagger \\
        \mathbf{g} & \mathcal H \\
    \end{pmatrix}
    \begin{pmatrix}
        1 \\
        \mathbf{x} \\
    \end{pmatrix}
=
\epsilon
\begin{pmatrix}
1 & 0 \\
0 & \mathbf{S} \\
\end{pmatrix}
\begin{pmatrix}
        1 \\
        \mathbf{x} \\
\end{pmatrix} \label{scieq}
\end{equation}
for determining the orbital-rotation parameters $\mathbf{x}$. 
The matrix elements of the overlap matrix \(\mathbf{S}\) are given by
\begin{equation}
    \begin{aligned}
     S_{pq,rs} &\equiv \langle \Psi_q^p|\Psi^r_s\rangle, \\
        S_{ai,bj} &= \delta_{ij}\delta_{ab},\\
        S_{ti,uj} &= \delta_{ij}\delta_{tu} - \delta_{ij}\gamma_{ut},\\
        S_{at,bu} &= \delta_{ab}\gamma_{tu},\\
    \end{aligned}
\end{equation}
in which the one-particle reduced density matrix $\gamma$ is given by
\begin{eqnarray}
    \gamma_{tu} &=& \langle \Psi_0|a_t^\dagger a_u|\Psi_0\rangle.
\end{eqnarray}
The overlap matrix becomes diagonal when the natural orbitals within the CAS space are employed.
Note that the stationary conditions $\frac{\partial E^{\text{CASSCF,(2)}}_{\text{SCI}}(\mathbf{x})}{\partial x_{pq}}$ give the corresponding equation to determine $x^\ast$, which is the complex conjugate of Eq. (\ref{scieq}). 

The alternate solution of Eq. (\ref{scieq}) to determine $\mathbf{x}$ to update orbitals and the FCI equation in the active space to update $C_I$'s, when converged, leads to a vanishing $\mathbf{g}$ vector. 
Importantly, in spite of the approximation introduced in the SCI scheme, $\mathbf{g}$ is equal to the ``orbital gradient vector'', the gradient of the CASSCF energy with respect to the orbital-rotation parameters.
Therefore, the CASSCF wave function obtained using the SCI scheme 
correctly satisfies the stationary condition for orbital rotation, namely, it satisfies the generalized Brillouin theorem \cite{1968LevyIJQC}
\begin{eqnarray}
    \langle \Psi^{\text{CASSCF}}|[\hat H, a_p^\dagger a_q]|\Psi^{\text{CASSCF}}\rangle=0.
\end{eqnarray}

$\mathcal{H}$ is identical to the $\frac{\partial^2 E^{\text{CASSCF}}}{\partial x_{pq}^\ast \partial x_{rs}}$
block of the Hessian matrix.
The SCI equation for determining orbital-rotation parameters, Eq. (\ref{scieq}), 
is equivalent to the neglect of 
the $\frac{\partial^2 E^{\text{CASSCF}}}{\partial x_{pq} \partial x_{rs}}$
and $\frac{\partial^2 E^{\text{CASSCF}}}{\partial x^\ast_{pq}\partial x^\ast_{rs}}$
blocks of the Hessian matrix in the second-order CASSCF algorithm. \cite{2018ReynoldsTJoCP}
Note that the presence of the $\frac{\partial^2 E^{\text{CASSCF}}}{\partial x_{pq} \partial x_{rs}}$
and $\frac{\partial^2 E^{\text{CASSCF}}}{\partial x^\ast_{pq}\partial x^\ast_{rs}}$
blocks of the Hessian matrix couples the equations for $x$ and $x^\ast$ in the second-order CASSCF algorithm.
In the more approximate SCI scheme, the equations for $x$ and $x^\ast$ are uncoupled; they are simply the complex conjugate of each other. 
The matrix elements for the nonzero blocks of $\mathbf{g}$, the $ai$, $at$, and $ti$ blocks,
can be written in a computationally convenient form as 
\begin{equation}
    \begin{aligned}
        g_{ai} &= F_{ai}^{\text{eff}}, \\ 
        g_{at} &= \sum_u F_{tu}^{\text{core}}\gamma_{ut} + Q_{at}, \\
        g_{ti} &= F_{ti}^{\text{eff}} 
        -\sum_u F_{ui}^{\text{core}}\gamma_{tu} - Q_{it}^\ast.
    \end{aligned}
\end{equation}
The ``effective'' Fock matrix $F^{\text{eff}}$ is defined as the sum of the ``core'' Fock matrix $F^{\text{core}}$ and a contribution from the active space 
\begin{eqnarray}
F^{\text{eff}}_{pq} &=& F^{\text{core}}_{pq}+\sum_{tu} [(pq|tu)-(pu|tq)]\gamma_{tu}, \\
    F^{\text{core}}_{pq} &=& h_{pq}+\sum_i [(pq|ii)-(pi|iq)], 
\end{eqnarray}
and an auxiliary $\mathbf{Q}$ matrix is defined as
\begin{equation}
   Q_{pt}=\sum_{uvw}(pu|vw)\Gamma_{tuvw}
\end{equation}
using two-particle reduced density matrix
\begin{eqnarray}
    \Gamma_{tuvw} &=& \langle \Psi_0|a_t^\dagger a_v^\dagger a_w a_u|\Psi_0\rangle.
\end{eqnarray}

The full evaluation of $\mathcal{H}$ requires up to three-particle reduced density matrix in the active space
and is computationally expensive.\cite{1980SiegbahnPS}
To improve computational efficiency, 
we have adopted an approximation of $\hat{H}$ in $\mathcal{H}$ using an effective Fock operator $\hat{F}^\mathrm{eff}$\cite{1980RoosCP,1980RoosIJQC}
\begin{equation}
\begin{aligned}
\hat F^{\mathrm{eff}} &= \sum_{pq} F^{\mathrm{eff}}_{pq}\, a^\dag_p a_q. 
\end{aligned}
\end{equation}
The simplified matrix elements of \(\mathcal{H}\) are then given by
\begin{equation}
    \begin{aligned}
        \mathcal{H}_{ai,bj} &=\delta_{ij}F^{\text{eff}}_{ab}-\delta_{ab}F^{\text{eff}}_{ji},\\
        \mathcal{H}_{ai,uj} &=\delta_{ij}F^{\text{eff}}_{au}-F^{\text{eff}}_{av}\gamma_{uv},\\
        \mathcal{H}_{ai,bu} &=-\delta_{ab}F^{\text{eff}}_{ti}\gamma_{tu},\\
        \mathcal{H}_{ti,uj} &=F^{\text{eff}}_{ji}\gamma_{ut}+\delta_{ij}\left[ F^{\text{eff}}_{tu}-F^{\text{eff}}_{tv}\gamma_{uv}-F^{\text{eff}}_{vu}\gamma_{vt}-\sum_{vw}(\Gamma_{tuvw}-\gamma_{tu}\gamma_{vw})F_{vw}\right],\\
        \mathcal{H}_{ti,bu} &=0,\\
        \mathcal{H}_{at,bu} &=\gamma_{tu}F^{\text{eff}}_{ab}+\delta_{ab}\sum_{vw}F^{\text{eff}}_{vw}(\Gamma_{tu,vw}-\gamma_{tu}\gamma_{vw}).\\        
    \end{aligned}
\end{equation}
The cost 
of acting this approximate \(\mathcal{H}\) on a trial vector is essentially negligible.
We mention that $\hat{H}$ in $\mathcal{H}$ may also be approximated using the ``Dyall Hamiltonian'', \cite{Dyall95} which accounts for the complete contribution from the Hamiltonian within the active space. \cite{2019KollmarJCC}

As discussed above, the SCI scheme can be viewed as an approximate form of the second-order algorithm.
The SCI formulation presented here may also be derived by neglecting the $\frac{\partial^2 E^{\text{CASSCF}}}{\partial x_{pq} \partial x_{rs}}$
and $\frac{\partial^2 E^{\text{CASSCF}}}{\partial x^\ast_{pq}\partial x^\ast_{rs}}$
blocks of the Hessian matrix in the second-order CASSCF algorithm \cite{2015BatesJCP} and further approximating
the Hamiltonian operator in the $\frac{\partial^2 E^{\text{CASSCF}}}{\partial x^\ast_{pq} \partial x_{rs}}$ block with an effective Fock operator. The SCI scheme maintains the exact orbital gradient vector. Therefore, upon convergence, the SCI scheme yields the same result as the second-order algorithm. The SCI calculations are expected to converge more slowly than the second-order algorithm, because of the approximations for the Hessian matrix. However, these approximations dramatically reduce the computational cost for each iteration.
The SCI scheme thus 
has the potential to compute much larger molecules.

\subsubsection{The CD-based implementation}

We now discuss the computational aspects of the current implementation.
In the solution of the SCI equation [Eq. (\ref{scieq})] with the traditional algorithm, the construction of the auxiliary matrix \(\mathbf{Q}\) dominates the computational cost. 
The evaluation of \(\mathbf{Q}\) involves the calculation of the two-electron integrals of the form \((pu|vw)\), 
with a formal scaling of \(\mathcal{O}(N_{\mathrm{act}}N_{\mathrm{ao}}^4)\). Here \(N_{\mathrm{act}}\) denotes the number of active orbitals and \(N_{\mathrm{ao}}\) the number of atomic orbitals.
For comparison, the second most expensive step is the construction of the effective Fock matrix $F^{\mathrm{eff}}$ 
which exhibits a formal scaling of \(\mathcal{O}(N_{\mathrm{ao}}^4)\).

The use of Cholesky decomposition of the two-electron integrals or the closely related density fitting technique 
can significantly reduce the computational cost 
associated with integral transformation in the CASSCF calculations.
\cite{1996Ten-noJCP,2008AquilanteJCP,2015BatesJCP,2018ReynoldsJCP,2020KreplinJCP,2021NottoliJCTC,WangC,Evangelista24}
Here we focus on the CD-based calculations of the auxiliary matrix \(\mathbf{Q}\) and the Fock matrices in
the X2C-CASSCF calculations.
The Cholesky decomposition of two-electron AO integrals 
\begin{equation}
    (\mu\nu|\sigma\lambda) = \sum_P^{N_\mathrm{cd}} L^P_{\mu\nu}L^P_{\sigma\lambda}~,~ N_{\mathrm{cd}} = M N_{\mathrm{ao}},
\end{equation}
condenses the information in the two-electron integral matrix $(\mu\nu|\sigma\lambda)$ into \(N_{\mathrm{cd}}\) Cholesky vectors \(L_{\mu\nu}\).
We introduce $M$ as the ratio between \(N_{\mathrm{cd}} \) and \(N_{\mathrm{ao}}\). 
\(M\) typically ranges from 3 to 10 depending on the threshold of Cholesky decomposition.\cite{2024PedersenWCMS} 
The total size of the Cholesky vectors amounts to around $\frac{1}{2}MN_{\mathrm{ao}}^3$,
much smaller than the size of $\frac{1}{8}N_{\mathrm{ao}}^4$ for the two-electron integral matrix.

Using the Cholesky vectors, the auxiliary matrix \(\mathbf{Q}\) is obtained as
\begin{equation}
    Q_{pt} = \sum_{Pu}L^P_{pu}\tilde L^P_{tu}, \label{QCD}
\end{equation}
in which $L^P_{pu}$ is a Cholesky vector in the MO representation
\begin{equation}
    L^P_{pt}=\sum_{\mu\nu} C_{\mu p}^\ast  L^P_{\mu\nu} C_{\nu t} \label{CDtran}
\end{equation}
and $\tilde L^P_{tu}$ is obtained by contracting an active-active block of Cholesky vector, \(L_{tu}^P\), with the active space two-particle reduced density matrix,
\begin{equation}
    \tilde L^P_{tu} = \sum_{vw} L^P_{vw} \Gamma_{tuvw}. \label{Ltilde}
\end{equation}
The transformation of the Cholesky vectors from the AO to MO representation in Eq. (\ref{CDtran}) has a formal scaling of \(\mathcal{O}(MN_{\mathrm{act}}N_{\mathrm{ao}}^3)\), significantly reduced compared to the scaling of \(\mathcal{O}(N_{\mathrm{act}}N_{\mathrm{ao}}^4)\) for the integral transformation using the traditional algorithm.
The other two steps in Eqs. (\ref{QCD}) and (\ref{Ltilde}) have lower computational cost, with formal scalings of 
\(\mathcal{O}(MN_{\mathrm{MO}}N_{\mathrm{act}}^2N_{\mathrm{ao}})\) and \(\mathcal{O}(MN_{\mathrm{act}}^4N_{\mathrm{ao}})\), respectively. In two-component CASSCF calculations, the number of total molecular orbitals ($N_{\mathrm{MO}}$) is around twice $N_{\mathrm{ao}}$, while the number of active orbitals $N_{\mathrm{act}}$ is much smaller than $N_{\mathrm{ao}}$.

The calculation of the Fock matrices $F^{\mathrm{core}}$ and $F^{\mathrm{eff}}$ using Cholesky vectors 
exhibits a formal scaling of \(\mathcal{O}(M(N_{\mathrm{core}}+N_{\mathrm{act}})N_{\mathrm{ao}}^3)\). 
The number of ``core'' orbitals $N_{\mathrm{core}}$ is usually much larger than the number of active orbitals $N_{\mathrm{act}}$. Therefore, the cost of constructing $F^{\text{core}}$ and $F^{\text{eff}}$ is higher than the calculation of the auxiliary matrix \(\mathbf{Q}\) and dominates the calculations of orbital-rotation
parameters in CD-based CASSCF calculations. 
This also indicates that, with the CD-based implementation and the use of the SCI scheme, the computational cost for the calculation of the orbital-rotation parameters in each X2C-CASSCF iteration is comparable to that in each X2C Hartree-Fock iteration. Therefore, the current X2C-CASSCF implementation are applicable to calculations of 
sizable molecules.

\section{Computational details}
\label{sec:comp_details} 

We have implemented the CD-based relativistic exact two-component CASSCF method in a locally modified version of the PySCF program package\cite{2018SunWCMS,2020SunJCP}. All X2C-CASSCF calculations presented here have used this implementation. We have used the implementation of the X2CMP scheme \cite{Wang25a} and the current implementation of the X2Ccorr scheme in the PySCF program for all the X2CMP and X2Ccorr calculations presented here. Both X2CMP and X2Ccorr schemes have also been implemented in the CFOUR program package \cite{CFOURfull,Matthews20a} for the verification of the implementation. The X2CAMF program \cite{2022ZHANG} has been used to provide the X2CAMF integrals as well as the atomic density matrices for the construction of the X2CMP integrals. 
In the calculations of the quantum electrodynamics (QED) effects, we augment the four-component one-electron Hamiltonian used in the X2CMP scheme with QED integrals to account for the contributions from vacuum polarization \cite{uehlingPolarizationEffectsPositron1935, fullertonAccurateEfficientMethods1976} and self energy \cite{flambaumRadiativePotentialCalculations2005}. 
The QED integrals have been calculated using the CFOUR program package and imported into the PySCF program for the calculations of QED corrections. 
A tight threshold of \(1\times10^{-7}\) Hartree for Cholesky decomposition of two-electron integrals has been used for all calculations. 

We have performed benchmark exact two-component state averaged CASSCF calculations of the zero-field splittings in the ground $X^3\Sigma^-$ states of chalcogen diatomics, including
\ce{O2}, \ce{SO}, \ce{S2}, \ce{SeO}, \ce{SeS}, \ce{Se2}, \ce{TeO},
\ce{TeS}, \ce{TeSe}, and \ce{Te2}. 
All calculations have used active spaces comprising 
8 electrons in 12 spinors to include all valence $np$ electrons
and orbitals in the active spaces.
Each CASSCF calculation has averaged over the lowest three electronic states
corresponding to the $0^{+}$ and $1_{g}$ components of the $^3\Sigma^-$ state. 
The uncontracted aug-cc-pVTZ basis sets have been used for O \cite{Kendall92} and S \cite{Woon93}, while the Dyall's aVTZ basis sets have been used for Se and
Te \cite{Dyall02_dyalltz_4p5p6p,Dyall06_dyalltz_revise_4p5p6p}. The experimental bond lengths \cite{1979HuberMSaMSICoDM}, 1.20752 \AA~for \ce{O2}, 1.481 \AA~for \ce{SO}, 1.889 \AA~for \ce{S2}, 1.648 \AA~for {SeO}, 2.037 \AA~for \ce{SeS}, 2.166 \AA~for \ce{Se2}, 1.825 \AA~for \ce{TeO}, 2.230 \AA~for \ce{TeS}, 2.372 \AA~for \ce{TeSe} and 2.557 \AA~for \ce{Te2} have been used for these molecules. 

The X2CAMF schemes based on the Dirac-Coulomb (DC), Dirac-Coulomb-Gaunt (DCG), and Dirac-Coulomb-Breit (DCB) Hamiltonians
have been used to compare with the corresponding X2C-1e calculations to demonstrate the contributions from the two-electron spin-same-orbit interaction, two-electron spin-other-orbit interaction, and the gauge term.  
The comparison between X2CMP and X2CAMF calculations based on the DCG Hamiltonian demonstrates the contributions from the scalar-relativistic two-electron picture change and from the two- and three-center relativistic two-electron integrals. 
The X2Ccorr scheme has been used to include the picture-change correction to the fluctuation potential, which includes the electron spin-spin coupling of importance to the calculations of zero-field splittings. 
The QED corrections have also been included for comparison with the other corrections.




To further demonstrate the applicability of the Cholesky decomposition based implementation of exact two-component CASSCF method using the SCI scheme, 
we have calculated the spin-orbit and ligand-field splittings of the Nd $4f^3$ configurations 
in the \ce{Nd^{3+}} aquo-ions. 
All these calculations of \ce{Nd^{3+}} aquo-ions have used the same active space as used in Ref. \citenum{2025NielsenIC} that comprises 14 Nd $4f$ orbitals and 3 open-shell electrons.
We follow Nielson \textit{et al.}\cite{2025NielsenIC} in the construction of the ligand field for the aquo-ions by using point charges or with explicit inclusion of water molecules. 
First, benchmark studies of the basis-set effects and the relativistic two-electron contributions have been carried out using a model system, in which the \ce{Nd^{3+}} free ion is embedded in  point charges with cubic symmetry, using 8 point charges of -1.0 $e$ placed 2.333 {\AA}  from the \ce{Nd^{3+}}
center. 
The uncontracted Dyall's VTZ and VQZ (VTZ-unc and VQZ-unc)  basis sets\cite{2010GomesTCA} as well as
the uncontracted ANO-RCC (ANO-RCC-unc) basis sets \cite{2008RoosJPCA} 
have been used in the study of basis-set effects. Calculations with spin-orbit contracted VTZ (VTZ-SO) basis set using the primitive functions of the VTZ-unc basis set have also been carried out to test the accuracy of the new SO contraction scheme. \cite{2024ZhangTJoCP} 
The X2C-1e, X2CAMF(DC), X2CAMF(DCG), X2CAMF(DCB),
X2CMP(DCB), and X2Ccorr calculations have been performed to study relativistic two-electron contributions. The QED corrections have also been computed to understand the relative importance with respect to the relativistic two-electron contributions.

We have further computed \ce{Nd^{3+}} aquo-ions with explicit inclusion of the first and second coordination shells of water molecules.  
As discussed in Ref. \citenum{2025NielsenIC}, the first coordination shells most probably take coordination numbers 8 or 9. 
We have computed the corresponding model systems \ce{[Nd(H2O)8]^3+} and \ce{[Nd(H2O)9]^3+}. 
The addition of a second coordination shell to the models with coordination numbers 8 and 9 gives rise to model systems with 23 and 26 water molecules, \ce{[Nd(H2O)23]^3+} and \ce{[Nd(H2O)26]^3+}, respectively. 
We have also studied two additional structures [Nd(H$_2$O)$_8^\ast$]$^{3+}$, [Nd(H$_2$O)$_9^\ast$]$^{3+}$ 
obtained by simply removing the
water molecules in the second coordination shell in \ce{[Nd(H2O)23]^3+} and \ce{[Nd(H2O)26]^3+}, namely, including the effects of the second coordination shell on the structures of the first coordination shell.
We have used the structures for these aquo-ions optimized in Ref. \citenum{2025NielsenIC} using density functional theory calculations. 
The VTZ-SO basis set for Nd and the SFX2C-1e recontracted cc-pVDZ basis sets for O and H \cite{CFOUR_Recontracted_cc_basis} have been used in all calculations of the \ce{Nd^{3+}} aquo-ions.

\section{Results and Discussions}

\label{sec:results}

\subsection{Zero-field splitting of chalcogen diatomics}

\label{subsec:chalcogen} 

Computed zero-field splittings of the ground $X^3\Sigma^-$ states of chalcogen diatomic molecules, namely,
the splittings between the $0^{+}$ and $1_{g}$ components of the $X^3\Sigma^-$ states
are summarized
in Table \ref{tab:zfs_ham}.
The three columns under $\Delta\text{X2CAMF}$ represent the relativistic two-electron contributions.
The ``Coulomb'' column summarizes the difference between the X2CAMF(DC) and X2C-1e results, 
which are the relativistic two-electron contributions from the Coulomb interaction. 
The ``Gaunt'' column represents the contributions from the Gaunt term, given as the differences between the X2CAMF(DCG) and X2CAMF(DC) results.
The contributions from the gauge term are calculated as the difference between the X2CAMF(DCB) and X2CAMF(DCG) results and summarized in the ``gauge'' column of Table \ref{tab:zfs_ham}. 
 As expected, the X2C-1e scheme with only the one-electron spin-orbit term overestimates
 the zero-field splittings. 
The relativistic two-electron interactions in the Coulomb and Gaunt terms
``screen'' the one-electron spin-orbit term.
As expected, the contributions from the spin-same-orbit interaction in the Coulomb term are relatively more significant for light elements. They reduce the zero-field splittings in O$_2$, S$_2$, Se$_2$, and Te$_2$ by 50\%, 25\%, 15\%, and 7\%, respectively. The contributions from the spin-other-orbit interaction in the Gaunt term are around 1/4 those of the Coulomb term for all molecules studied here. 
The contributions from the gauge term are much smaller than those from the Coulomb or Gaunt term. 
They are almost negligible; the largest contribution appears in the case of Te$_2$ and amounts to only 0.3 cm$^{-1}$.

\begin{table}[ht]
\centering \caption{Zero-field splittings (\wavenumber) in the ground $X^{3}\Sigma^-$ states of chalcogen
diatomic molecules obtained from state-averaged CASSCF calculations for the three components of the $X^3\Sigma^-$ states. 
``Coulomb'', ``Gaunt'', and ``gauge'' represents the relativistic two-electron contributions from the Coulomb interaction, the Gaunt term, and the gauge term. ``$\Delta$X2CMP'' includes scalar two-electron picture-change correction and the contributions from multi-center relativistic two-electron integrals. ``$\Delta$QED'' denotes the QED corrections. ``$\Delta$X2Ccorr'' is the correction due to the picture change of fluctuation potential.}
\label{tab:zfs_ham} %
\setlength{\tabcolsep}{2pt}
\begin{tabular}{@{}cccccccccc@{}}
\hline 
 & ~ ~X2C-1e ~ ~ & \multicolumn{3}{c}{$\Delta$X2CAMF$^{a}$}  & ~ $\Delta$X2CMP$^{b}$ ~ & ~ $\Delta$QED$^{c}$ ~ & ~ $\Delta$X2Ccorr$^{d}$~  & Total & \tabularnewline
 \hline
  &         & Coulomb  & Gaunt  & gauge  &                       &                & & & \tabularnewline
\hline
\ce{O2}  & 6.9  & -3.3  & -0.7  & 0.00  & -0.2  & -0.1 & 0.9 & 3.6 &\tabularnewline
\ce{SO}  & 19.6  & -6.6  & -1.6  & 0.00  & -0.2  & -0.1 & 0.2 & 11.6 &\tabularnewline
\ce{S2}  & 38.6  & -10.8  & -2.5  & 0.00  & -0.2  & -0.0 & 0.1 & 25.4 & \tabularnewline
\ce{SeO}  & 246.6  & -43.8  & -10.1  & 0.01  & -1.0  & 0.2 & -0.3 & 192.7 & \tabularnewline
\ce{SeS}  & 263.4  & -44.8  & -10.0  & 0.00  & -0.8  & 0.4 & -0.1 & 208.8 & \tabularnewline
\ce{Se2}  & 646.0  & -87.1  & -19.4  & -0.03  & -1.7  & 1.0 & -0.2 & 540.2 & \tabularnewline
\ce{TeO}  & 999.9  & -102.6  & -24.8  & 0.09  & -2.3  & 2.1 & -0.8 & 873.7 & \tabularnewline
\ce{TeS}  & 937.0  & -96.1  & -22.6  & -0.02  & -1.8  & 1.9 & -0.3 & 819.9 & \tabularnewline
\ce{TeSe} & 1437.0  & -132.5  & -31.2  & 0.12  & -2.6  & 2.4 & -0.3 & 1275.5 & \tabularnewline
\ce{Te2}  & 2249.3  & -159.3  & -39.3  & 0.27  & -3.6  & 3.3 & -0.3 & 2053.4 & \tabularnewline
\hline 
\end{tabular}
\medskip
\parbox{\linewidth}{\footnotesize
${^a}$ The differences between X2CAMF and X2C-1e results.\\
$^{b}$ the difference between X2CMP(DCB) and X2CAMF(DCB) results.\\
$^{c}$ the difference between X2CMP(DCB)+QED and X2CMP(DCB) results.\\
$^{d}$ the difference between X2Ccorr(DCG) and X2CMP(DCG) results.
}
\end{table}

The column ``$\Delta$X2CMP'' gives the differences between the X2CMP(DCG) and X2CAMF(DCG) results, which include the contributions from the scalar two-electron picture-change effects and those from two- and three-center relativistic two-electron integrals. 
These contributions are generally small. 
They are less than 2\% of the total X2CMP values, except for O$_2$.
Although it amounts to around 5\% of the X2CMP value in O$_2$, the absolute value of 0.2 {\wavenumber} for this contribution remains small. 
We also emphasize that, as shown in Table \ref{tab:zfs_ham} the QED corrections are of similar magnitude and often have an opposite sign compared to the differences between the X2CMP and X2CAMF results. Therefore, in order to improve the X2CAMF calculations, it seems necessary to augment the X2CMP calculations with QED corrections. This observation is consistent with the findings in the study of core-level spectroscopy. \cite{Wang25, Southworth15} 

The contributions from the picture-change correction to the fluctuation potential, the results in the $\Delta$X2Ccorr column, are generally small for these zero-field splittings, except for O$_2$. The $\Delta$X2Ccorr value for O$_2$ amounts to 0.9 \wavenumber, about 25\% of the total value. This is consistent with the understanding that the electron spin-spin coupling makes significant contributions to zero-field splittings in light elements. The SSC contributions decay with the increase of the nuclear charge. Nevertheless, this underlies the importance of including the picture-change correction for fluctuation potential to obtain a unified accuracy across the periodic table. 


\begin{table}[ht]
\centering \caption{Zero-field splittings (\wavenumber) in the ground $X^{3}\Sigma^-$ states of chalcogen
diatomic molecules. 
The three components of the $X^{3}\Sigma^-$ states have been included in the state-averaged X2C-CASSCF calculations. The state-interaction CASPT2-SO and four-component intermediate Hamiltonian Fock-space coupled-cluster (IHFSCC) results are taken from Ref. \citenum{2011RotaJCP} for comparison. }
\label{tab:zfs_methods} %
\begin{tabular}{lcccc}
\hline 
 & Expt.\cite{1979HuberMSaMSICoDM}  & X2C-CASSCF   & CASPT2-SO \cite{2011RotaJCP} & IHFSCC \cite{2011RotaJCP}\tabularnewline
\hline 
\ce{O2}  & 4.0  & 3.6  & -- & -- \tabularnewline
\ce{SO}  & 10.5  & 11.6  & -- & -- \tabularnewline
\ce{S2}   & 23.5  & 25.4  & 19  & 25 \tabularnewline
\ce{SeO}  & 166  & 192.7  & 138  & 137 \tabularnewline
\ce{SeS}& 205  & 208.8 & 165  & 192 \tabularnewline
\ce{Se2}& 510  & 540.2  & 446  & 574 \tabularnewline
\ce{TeO}  & 789$^a$ & 873.7  & 635  & 468 \tabularnewline
\ce{TeS}  & 836$^a$ & 819.9  & 675  & 619 \tabularnewline
\ce{TeSe}  & 1233$^b$ & 1275.5  & 1073  & 1238 \tabularnewline
\ce{Te2}  & 1975  & 2053.4  & 1793  & 2178 \tabularnewline
\hline 

\end{tabular}
\medskip
\parbox{\linewidth}{\footnotesize
${^a}$ Experimental values from Ref.~\citenum{Winter1982TeO}\\
$^{b}$ Experimental value from Ref.~\citenum{1989FinkJoMS}.
}
\end{table}
As shown in 
Table \ref{tab:zfs_methods},
the final X2C-CASSCF results compare favorably with the measured values, with errors around or less than 10\% of the total values.  
The CASPT2-SO results generally underestimate these splittings. 
While the current X2C-CASSCF calculations perform state averaging for the three target states, the CASPT2-SO calculations include three singlet states and one triplet states in the active space and perform state averaging for these six states. This accounts for the lower accuracy of the CASPT2-SO results. The four-component IHFSCC calculations produce as accurate results as the X2C-CASSCF calculations for S$_2$, SeO, SeS, Se$_2$, TeSe, and Te$_2$. However, the corresponding results for TeO and TeS are obviously less accurate. 
This can be attributed to large wave function relaxation effects due to the addition of two electrons to closed-shell reference states in these IHFSCC calculations, which are often difficult to treat accurately in the CCSD framework. 

\subsection{Spin-orbit and ligand-field splittings in Nd aquo-ions}

The solvation models for the lanthanide ions are of topical interest to lanthanide chemistry.\cite{1968CarnallJCP,1979HabenschussJCP,1979HabenschussJCPa,1980HabenschussJCP,1995KowallJACS,2012DAngeloC-EJ,2014ZhangIC,2021ShieryIC,2025NielsenNC}
Here we demonstrate the applicability of the current CD-based implementation of the X2C-CASSCF method
using calculations of spin-orbit and ligand-field splittings in Nd aquo-ions. 
The three open-shell $4f$ electrons in the free \ce{Nd^{3+}} ion
take $^4I$ as the lowest manifold of electronic states.
Spin-orbit coupling splits the $^4I$ manifold into
four sub-levels, the $^4I_{9/2}$, $^4I_{11/2}$, $^4I_{13/2}$ and $^4I_{15/2}$ states. 
The magnitude of the spin-orbit splittings amounts to several thousand \wavenumber. The ligand field created by the solvent molecules further breaks the degeneracy within the sub-levels. 
The ligand-field splittings span a few hundred \wavenumber and are fingerprint information about the structures of the aquo-ions.
S{\o}rensen and collaborators have recently significantly improved the precision for the measured ligand-field splittings using high resolution spectroscopy. \cite{2021KofodIC,2025NielsenIC,2025NielsenNC}
In particular, the measured ligand-field splittings for the $^4I_{9/2}$ and $^4I_{13/2}$ manifolds provide valuable benchmark results for testing computational methods. 

We have performed X2C-CASSCF calculations with state averaging over the 52 electronic states in the $I_4$ manifold
to study the relativistic two-electron contributions 
as well as the contributions from the structures of the aquo-ions to these energy splittings. 
We benchmark the X2C-CASSCF calculations using the measured splittings. The comparison with the computed splittings using the state-interaction CASSCF (CASSCF-SO) calculations \cite{2025NielsenIC} also serves as a validation of the current implementation and calculations. 
Here we expect the X2C-CASSCF ligand-field splittings to be similar to the CASSCF-SO results using
the same active space, since the ligand-field splittings do not directly originate from relativistic effects and are expected to be less sensitive to the treatment of relativistic effects. In contrast, we expect the X2C-CASSCF spin-orbit splittings  
to differ more substantially from the CASSCF-SO results, because of the inclusion of higher-order spin-orbit effects. 

\begin{table}[ht]
\caption{Center of gravity levels (cm$^{-1}$) of \ce{Nd^{3+}} embedded in a cubic point-charge
field obtained from the X2CAMF(DCB)-CASSCF calculations averaging over the lowest 52 electronic states. }
\label{tab:basis} 
\begin{tabular}{ccccc}
\hline 
 & VTZ-SO  & VTZ-unc  & VQZ-unc & ANO-RCC-unc\tabularnewline
 \hline
$^{4}I_{9/2}$  & 73  & 61  & 63  & 62\tabularnewline
$^{4}I_{11/2}$  & 1867  & 1839  & 1840  & 1835 \tabularnewline
$^{4}I_{13/2}$  & 3827  & 3782  & 3784  & 3775 \tabularnewline
$^{4}I_{15/2}$  & 5914  & 5854  & 5855  & 5843 \tabularnewline
\hline 
\end{tabular}
\end{table}

\subsubsection{Spin-orbit splittings in the \ce{Nd^{3+}} ion embedded in point charges}

We first focus on a model system with a \ce{Nd^{3+}} ion embedded in a set of point charges as constructed in Ref. \citenum{2025NielsenIC} to study basis-set effects and the relativistic two-electron contributions to the spin-orbit splittings.
We use the center of gravity levels, namely, the average energies of a sub-level. The energies relative to the energy of the lowest electronic states are summarized in Table \ref{tab:basis} and Table \ref{tab:nd_ham}.
As shown in Table \ref{tab:basis}, the computed spin-orbit splittings are insensitive to the basis sets. The VTZ-unc basis set provides nearly converged splittings; the differences between the results obtained using VTZ-unc and VQZ-unc basis sets are a few to 10 \wavenumber. The ANO-RCC-unc basis set also produces very similar splittings.
Meanwhile, the spin-orbit contraction works well in these calculations. The VTZ-SO basis set produces results with errors of 10-60 {\wavenumber} compared to the VTZ-unc basis set. 
This justifies the use of the spin-orbit contracted basis set for calculations of the aquo-ions. 

\begin{table}[ht]
\caption{Center of gravity levels (cm$^{-1}$) of \ce{Nd^{3+}} embedded in a cubic point-charge
field obtained from X2C-CASSCF calculations using Dyall's VTZ-unc basis set.
``Coulomb'', ``Gaunt'', and ``gauge'' represents the relativistic two-electron contributions from the Coulomb interaction, the Gaunt term, and the gauge term. 
``$\Delta$X2CMP'' includes scalar two-electron picture-change correction. ``$\Delta$QED'' denotes the QED corrections. ``$\Delta$X2Ccorr'' is the correction due to the picture change of fluctuation potential.}
\label{tab:nd_ham} %
\setlength{\tabcolsep}{2pt}
\begin{tabular}{@{}ccccccccc@{}}
\hline 
 & ~ ~X2C-1e ~ ~ & \multicolumn{3}{c}{$\Delta$X2CAMF$^{a}$} & ~ $\Delta$X2CMP$^{b}$ ~ & ~ $\Delta$QED$^{c}$ ~ & ~ $\Delta$X2Ccorr$^{d}$~ & Total \tabularnewline
 \hline
  &         & Coulomb  & Gaunt  & gauge  &                       &                & \tabularnewline
  \hline
$^{4}I_{9/2}$  & 59.8  & 2.0  & -0.5  & 0.1  & 0.1 & 0.0 & 0.0 & 61.4 \tabularnewline
$^{4}I_{11/2}$  & 4349.4  & -2356.6  & -153.5  & -0.8 & -0.4 & 10.5 & 1.6 & 1850.3 \tabularnewline
$^{4}I_{13/2}$  & 8654.0  & -4561.5 & -308.6  & -1.6 & -1.0  & 21.3 & 3.2 & 3805.8 \tabularnewline
$^{4}I_{15/2}$  & 12895.4  & -6578.3  & -461.0 & -2.5 & -1.6 & 31.8 & 4.9 & 5888.7 \tabularnewline
\hline 
\end{tabular}
\medskip
\parbox{\linewidth}{\footnotesize
${^a}$ The differences between X2CAMF and X2C-1e results.\\
$^{b}$ the difference between X2CMP(DCB) and X2CAMF(DCB) results.\\
$^{c}$ the difference between X2CMP(DCB)+QED and X2CMP(DCB) results.\\
$^{d}$ the difference between X2Ccorr(DCG) and X2CMP(DCG) results.
}
\end{table}

The relativistic two-electron contributions play a critical role in accurate calculations of the spin-orbit splittings in the $^4I$ manifold.
As shown in Table \ref{tab:nd_ham}, 
the contributions from the spin-same-orbit interaction in the Coulomb term reduce the corresponding X2C-1e values by more than 50\%. The contributions from the Gaunt term are smaller, but still account for nearly 10\% of the total values. 
Consistent with the observation in the literature \cite{Blume62,Blume63,1964BlumePR,Boettger00,Fedorov00}, the relativistic two-electron contributions appear to be more important in lanthanides than in main group elements.
The two-electron contribution to the spin-orbit splitting in Nd$^{3+}$ is even larger than the total value in terms of absolute magnitude. 
For comparison, the two-electron contribution accounts for only about 10\% of the total zero-field splitting in Te.

The gauge term only makes marginal contributions of up to a few \wavenumber. The differences between the X2CMP and X2CAMF results are also insignificant and amount to a few \wavenumber. This indicates that the scalar two-electron picture-change effects and the contributions from two- and three-center relativistic two-electron integrals are practically negligible in these calculations.
The QED corrections are of the order of tens of \wavenumber. They play a minor role in these calculations but are much larger than the contributions from the gauge term and scalar two-electron picture-change effects. The spin-orbit splittings of the $^4I$ manifold can be represented using the center of gravity levels of the $^4I_{11/2}$, $^4I_{13/2}$, $^4I_{15/2}$ states relative to that of the $^4I_{9/2}$ state. The final computational splittings of 1789 \wavenumber, 3744 \wavenumber, and 5827 {\wavenumber} are in good agreement with the corresponding measured values of 1882 \wavenumber, 3904 \wavenumber, and 5904 {\wavenumber} in aqueous solution. \cite{1968CarnallJCP}
We note that these splittings are not sensitive to the chemical environments; the values of 1897 \wavenumber, 3907 \wavenumber, and 5989 {\wavenumber} in gas phase \cite{2007WyartJPBAMOP}
and the values of 1882 \wavenumber, 3864 \wavenumber, and 5912 {\wavenumber} in crystal structures \cite{1961CarlsonJCP} agree closely with those in aqueous solution. The X2C-CASSCF splittings are about 10\% smaller than the corresponding CASSCF-SO values.
The differences can be attributed to higher-order relativistic corrections included in the X2C-CASSCF calculations. 

Based on these benchmark results, we choose to use the X2CAMF(DCB) scheme for further calculations of the aquo-ions. Note that the inclusion of the gauge term in the X2CAMF scheme only introduces negligible computational overhead.

\subsubsection{Ligand-field splittings in the \ce{Nd^{3+}} aquo-ions}

We present X2CAMF(DCB)-CASSCF results for the ligand-field splittings of the  $^{4}I{}_{9/2}$ and $^{4}I{}_{13/2}$ manifolds in Tables \ref{tab:nd_I9} and \ref{tab:nd_I13} for comparison with the recent measurement. 
The correlation coefficients ($r$) of the computed and measured energy levels are also reported as an additional parameter to describe the consistency of the computed and measured results.
These calculations have adopted the structures of aquo-ions optimized by Nielson \textit{et al.}  \cite{2025NielsenIC}.
As expected, the X2C-CASSCF ligand-field splittings agree very well with the corresponding CASSCF-SO results \cite{2025NielsenIC} using the same active space, with discrepancies no more than several \wavenumber. This close agreement validates the current X2C-CASSCF implementation and also supports the findings in Ref.~\citenum{2025NielsenIC} about the structures of the \ce{Nd^{3+}} aquo-ions.

Let us first focus on the aquo-ions containing the first coordinate shell with coordination numbers of 8 and 9, \ce{[Nd(H2O)_8]^{3+}} and \ce{[Nd(H2O)_9]^{3+}}.
The computed ligand-field splittings in \ce{[Nd(H2O)_9]^{3+}}
exhibit better correlation with the measured values. However, this model significantly underestimates the splitting between the fourth and fifth states in the $^{4}I{}_{9/2}$ manifold and that between the fifth and sixth states in the $^{4}I{}_{13/2}$ manifold. The structures \ce{[Nd(H2O)_23]^{3+}} and  \ce{[Nd(H2O)_26]^{3+}} 
include a second solvation shell. 
The inclusion of the second solvation shell 
improves the description for the aquo-ions by 
accounting for the interaction between the first and second coordination shells. 
The ligand-field splittings computed using \ce{[Nd(H2O)_26]^{3+}} agree quite well with the measured values, and represent a significant improvement compared to 
\ce{[Nd(H2O)_9]^{3+}}. In contrast, the results obtained using \ce{[Nd(H2O)_23]^{3+}} do not exhibit an significant improvement over those obtained using \ce{[Nd(H2O)_8]^{3+}}.
Therefore, the X2C-CASSCF calculations support the conclusion in Ref.~\citenum{2025NielsenIC} that the \ce{Nd^{3+}} aquo-ions favors the coordination number of 9.


\begin{table}[ht]
\caption{The X2CAMF(DCB)-CASSCF ligand-field splittings of the $^{4}I_{9/2}$ manifold of the \ce{[Nd(H2O)_n]^{3+}} aquo-ions. The energies (\wavenumber) of Kramers doublets relative to the lowest states in this manifold are reported. All X2C-CASSCF calculations have averaged over the lowest 52 states and have used spin-orbit contracted VTZ basis for Nd and SFX2C-1e contracted cc-pVDZ basis sets for O and H. 
``$r$'' in the last row represents the correlation coefficient with the
measured energy levels.}
\label{tab:nd_I9}
\setlength{\tabcolsep}{2pt}
\begin{tabular}{@{}cccccccc@{}}
\hline 
 State \# & Expt.  & \ce{[Nd(H2O)_8]^{3+}}  & $[\mathrm{Nd(H_2O)_{8}^{\ast}}]^{3+}$ & \ce{[Nd(H2O)_{23}]^{3+}} & \ce{[Nd(H2O)_{9}]^{3+}}  & $[\mathrm{Nd(H_2O)_{9}^\ast]^{3+}}$  & \ce{[Nd(H2O)_{26}]^{3+}}\tabularnewline
 \hline
2 & 121  & 55  & 69  & 74  & 119  & 107  & 103  \tabularnewline
3 & 181  & 237  & 178  & 186  & 143  & 152  & 150  \tabularnewline
4 & 318  & 240  & 209  & 216  & 282  & 251  & 248  \tabularnewline
5 & 370  & 284  & 282  & 287  & 293  & 316  & 314  \tabularnewline
$r$ & 1.000  & 0.915  & 0.969  & 0.968  & 0.993  & 0.998  & 0.998  \tabularnewline
\hline 
\end{tabular}
\end{table}


Since the direct interaction between the second coordination shell with the Nd$^{3+}$ ion is weak, the second coordination shell contributes mainly through the interaction with the first coordination shell.
\ce{[Nd(H2O)_8$^\ast$]^{3+}} and \ce{[Nd(H2O)_9$^\ast$]^{3+}} represent structures obtained by removing the water molecules in the second solvation shells of \ce{[Nd(H2O)_23]^{3+}} and \ce{[Nd(H2O)_26]^{3+}}. 
As shown in Tables \ref{tab:nd_I9} and \ref{tab:nd_I13}, the results obtained using \ce{[Nd(H2O)_8$^\ast$]^{3+}} and \ce{[Nd(H2O)_9$^\ast$]^{3+}} agree closely with those obtained using \ce{[Nd(H2O)_23]^{3+}} and \ce{[Nd(H2O)_26]^{3+}}. This verifies the intuition that the first coordination shell serves as ligands and the second coordination shell can be treated more approximately as implicit solvents that modify the structure of the first coordination shell. 

Although they appear to be consistent with the measured values, the X2C-CASSCF results for \ce{[Nd(H2O)_23]^{3+}} and \ce{[Nd(H2O)_26]^{3+}} underestimate the ligand-field splittings. It is of interest to further include dynamic-correlation contributions for further improvement of the computational results. 


\begin{table}[ht]
\caption{The X2CAMF(DCB)-CASSCF ligand-field splittings of the $^{4}I_{13/2}$ manifold of the \ce{[Nd(H2O)_n]^{3+}} aquo-ions. The energies (\wavenumber) of Kramers doublets relative to the lowest states in this manifold are reported. 
All X2C-CASSCF calculations have averaged over the lowest 52 states and have used spin-orbit contracted VTZ basis for Nd and SFX2C-1e contracted cc-pVDZ basis sets for O and H. 
``$r$'' in the last row represents the correlation coefficient with the
experimental energy levels.}
\label{tab:nd_I13} 
\setlength{\tabcolsep}{2pt}
\begin{tabular}{@{}ccccccccc@{}}
\hline 
 State \# & Expt.  & \ce{[Nd(H2O)_8]^{3+}}  & $[\mathrm{Nd(H2O)_{8}^{\ast}}]^{3+}$ & \ce{[Nd(H2O)_{23}]^{3+}} & \ce{[Nd(H2O)_{9}]^{3+}}  & $[\mathrm{Nd(H_2O)_{9}^\ast]^{3+}}$  & \ce{[Nd(H2O)_{26}]^{3+}}\tabularnewline
 \hline
2 & 34  & 8  & 28  & 35  & 47  & 28  & 19   \tabularnewline
3 & 74  & 67  & 52  & 52  & 72  & 68  & 59  \tabularnewline
4 & 95  & 73  & 90  & 94  & 89  & 85  & 75  \tabularnewline
5 & 127  & 124  & 135  & 140  & 112  & 114  & 107  \tabularnewline
6 & 162  & 179  & 159  & 161  & 134  & 138  & 134  \tabularnewline
7 & 203  & 217  & 169  & 184  & 180  & 177  & 168  \tabularnewline
$r$  & 1.000  & 0.990  & 0.945  & 0.961  & 0.994  & 0.999  & 0.999  \tabularnewline
\hline 
\end{tabular}
\end{table}

\section{Summary and outlook}
\label{sec:summary}

We report the development of a ``X2Ccorr'' scheme to include the picture-change correction to the fluctuation potential.
Importantly, the X2Ccorr scheme based on the Dirac-Coulomb-Gaunt Hamiltonian accounts for electron spin-spin coupling, which plays an important role
in accurate calculations of zero-field splittings and molecular fine structures. 
We also report a new implementation of relativistic two-component complete active space self consistent field (CASSCF) method using 
the Cholesky decomposition of two-electron integrals and the super-CI algorithms
for orbital optimization. 
Benchmark calculations for the zero-field splitting
of open-shell electronic states have been carried out with careful analysis on the contributions from
two-electron spin-orbit, Gaunt term, gauge term, quantum electrodynamics, and the picture-change correction to the fluctuation potential. 
We further demonstrate the applicability of the current CD-based CASSCF method using
calculations of spin-orbit and ligand-field splittings of the $^4I$ states of Nd aquo-ions with the explicit inclusion of up to the second coordination shells. 

Future development includes the extension of the current relativistic CASSCF implementation
to accommodate the use of restricted active space (RAS) to enable calculations of core-ionized and excited states.
We also plan to extend the X2Ccorr scheme to treat the contributions from the picture change of fluctuation potential to dynamic correlation, 
by using the first-order interacting space as the active space
\begin{eqnarray}
    \hat{H}^{\text{X2Ccorr}}&=& \hat{H}^{\text{X2CMMF}}+\frac{1}{4} \sum_{abij} [g^{\text{4c}}_{ab,ij}-g^{\text{nr}}_{ab,ij}] \{a_a^\dag a_b^\dag a_j a_i\}, \label{HX2Ccorr2}
\end{eqnarray}
in which $i, j, \cdots$ and $a, b, \cdots$ represent occupied and virtual orbitals, respectively. 
Since the picture-change correction for the fluctuation potential is expected to be small, we propose to add this correction to the X2CMMF scheme. 
The X2Ccorr Hamiltonian in Eq. (\ref{HX2Ccorr2}) can be used together with quantum-chemical methods
for treating dynamic correlation, including the coupled-cluster methods. 
Compared to an X2CMMF-CC calculation, an X2Ccorr-CC calculation using Eq. (\ref{HX2Ccorr2}) requires an additional partial four-component integral transformation 
to evaluate $g^{\text{4c}}_{ab,ij}$ and is compatible with the recent efficient implementation \cite{Liu18b,Zhang24} of X2C-CC methods using atomic-orbital-based algorithms and Cholesky decomposition of two-electron integrals.


\section{Acknowledgments}

The work 
was supported by the
National Science Foundation, under Grant No. PHY-2309253.
The computations have been carried out at
the Advanced Research Computing at Hopkins (ARCH) core
facility (rockfish.jhu.edu), which is supported by the NSF 
under Grant OAC-1920103.  

\bibliography{minimal}

\end{document}